\documentclass[a4paper,5p,10pt]{elsarticle}
\usepackage{graphicx,dcolumn,bm,amsmath,amssymb,mathtools,color,natbib}
\usepackage{tgtermes}
\usepackage[colorlinks]{hyperref}

\newcommand{\ud}{\mathrm{d}}
\newcommand{\ve}{\varepsilon}
\DeclareMathOperator{\sgn}{sgn}

\journal{Physics Letters B}

\biboptions{sort&compress}
\begin{document}
\begin{frontmatter}

\title{A new generic and structurally stable cosmological model without singularity\tnoteref{t1}
}
\tnotetext[t1]{This work is dedicated to memory of late Marek Szyd{\l}owski who passed away on the 8th of October 2020.}

\author{Orest Hrycyna}
\ead{orest.hrycyna@ncbj.gov.pl}
\address{Theoretical Physics Division, National Centre for Nuclear
Research, Ludwika Pasteura 7, 02-093 Warszawa, Poland}
 
\begin{abstract}
Dynamical systems methods are used to investigate a cosmological model with non-minimally coupled scalar field and asymptotically quadratic potential function. We found that for values of the non-minimal coupling constant parameter $\frac{3}{16}<\xi<\frac{1}{4}$ there exists an unstable asymptotic de Sitter state, free from a parallelly propagated singularity for $\frac{5}{24}\le\xi<\frac{1}{4}$,  giving rise to non-singular beginning of the universe. The energy density associated with this state depends on value of the non-minimal coupling constant and can be much smaller than the Planck energy density. For $\xi=\frac{1}{4}$ we found that the initial state is in form of the static Einstein universe. Proposed evolutional model, contrary to the seminal Starobinsky's model, do not depend on the specific choice of initial conditions in phase space, moreover, a small change in the model parameters do not change the evolution thus the model is generic and structurally stable. The values of the non-minimal coupling constant can indicate for a new fundamental symmetry in the gravitational theory. We show that Jordan frame and Einstein frame formulation of the theory are physically nonequivalent. 
\end{abstract}


\end{frontmatter}

\section{Introduction}

In theoretical cosmology we investigate future and past evolution of the Universe. Methodological tools used to explore dynamical behaviour are based on models derived from the general theory of relativity. The discovery of accelerated expansion of the Universe \cite{Riess:1998cb, Perlmutter:1998np} changed our understanding of the Universe and provoked growing interest in dynamical dark energy models \cite{Copeland:2006wr, Bahamonde:2017ize}. The so-called quintessence idea was formulated as the simplest cosmological model filled with a scalar field and a potential function in order to describe the current accelerated expansion of the Universe \cite{Peebles:1987ek, Ratra:1987rm, Wetterich:1987fm}.

The simples extension of the scalar field sector of a cosmological theory can be obtained by inclusion the non-minimal coupling term $-\xi R\phi^{2}$ between the gravity and the scalar field with the dimensionless parameter $\xi$ serving as the coupling constant \cite{Chernikov:1968zm, Callan:1970ze, Birrell:1979ip}. One can find various theoretical motivations for this term. Treating the general relativity as an effective theory such 
contribution naturally emerge in expansion \cite{Donoghue:1994dn}. Additionally, the 
non-minimal coupling between the curvature and the scalar field appears as a result of quantum corrections to the scalar field in curved space \cite{Birrell:1984ix, Parker:book} and is required by the renormalisation of theory \cite{Callan:1970ze}. 
The non-minimal coupling is also interesting in the context of 
superstring theory \cite{Maeda:1985bq} and induced gravity \cite{Accetta:1985du}.
From an effective theory approach the coupling constant $\xi$ becomes free parameter in the theory and should be obtained from some general considerations \cite{Muta:1991mw,Buchbinder:1992rb,Atkins:2010eq, Atkins:2010re}, from a more fundamental theory or its value should be estimated using the observational cosmological data \cite{Luo:2005ra, Nozari:2007eq, Szydlowski:2008zza, Atkins:2012yn, Hrycyna:2015vvs}.

The non-minimally coupled scalar field cosmology attracted a lot of attention by many authors over the years in the connection with an inflationary epoch and description of the current accelerated expansion of the universe \cite{Spokoiny:1984bd, Belinsky:1985zd, Amendola:1990nn, Faraoni:1996rf, Barvinsky:2008ia, Setare:2008pc, Uzan:1999ch, Amendola:1999qq, Holden:1999hm, Gannouji:2006jm, Carloni:2007eu, Bezrukov:2007ep, Kamenshchik:1995ib}. In connection with the standard model of particle physics the non-minimally coupled Higgs field plays also important role \cite{DeSimone:2008ei, Bezrukov:2008ej, Barvinsky:2009fy, Clark:2009dc}.

In the present paper, using dynamical systems methods, we investigate dynamical behaviour of the flat Friedmann-Robertson-Walker cosmological model filled with the non-minimally coupled scalar field. The potential is assumed as asymptotically quadratic function for large values of the scalar field. We investigate dynamical behaviour of the theory in original Jordan frame formulation as well as in conformally transformed Einstein frame. In the Jordan frame we find generic asymptotically unstable de Sitter state as the initial state for cosmological evolution. While in the Einstein frame this sate corresponds to asymptotically stable Einstein static universe. We show that both formulations of the theory are physically nonequivalent.

\section{The model}

The total action integral for the theory is composed of two parts
\begin{equation}
\label{eq:action}
S =S_{g} + S_{\phi}\,,
\end{equation}
where the gravitational interaction is described by the pure Einstein-Hilbert action
\begin{equation}
\label{eq:j_grav}
S_{g} = \frac{1}{2\kappa^{2}}\int\ud^{4}x\sqrt{-g}\,R \,,
\end{equation}
with the gravitational constant $\kappa^{2}=8\pi G$, and the matter part of the theory is given by the non-minimally coupled scalar field described by
\begin{equation}
\label{eq:j_m}
S_{\phi}=- \frac{1}{2}\int\ud^{4}x\sqrt{-g}\Big(\ve\nabla^{\alpha}\phi\,\nabla_{\alpha}\phi + \ve\xi R \phi^{2} + 2U(\phi)\Big)\,,
\end{equation}
where $\ve=\pm1$ corresponds to canonical and phantom scalar field, respectively. We work in units where $c=\hbar=1$.

Variation of the total action with respect to the metric tensor gives the field equations for the theory
$$
R_{\mu\nu}-\frac{1}{2}g_{\mu\nu}R=\kappa^{2}\,T^{(\phi)}_{\mu\nu}\,,
$$
where the energy-momentum tensor for the non-minimally coupled scalar field is given by
\begin{equation}
\begin{split}
T^{(\phi)}_{\mu\nu}  = &\,\, \ve \nabla_{\mu}\phi\nabla_{\nu}\phi -
\ve\frac{1}{2}g_{\mu\nu}\nabla^{\alpha}\phi\nabla_{\alpha}\phi -
U(\phi)g_{\mu\nu}  \\ & + \ve\xi\Big(R_{\mu\nu}-\frac{1}{2}g_{\mu\nu}R\Big)\phi^{2} +
\ve\xi\Big(g_{\mu\nu}\Box\phi^{2}-\nabla_{\mu}\nabla_{\nu}\phi^{2}\Big)\,.
\end{split}\nonumber
\end{equation}
The dynamical equation for the scalar field we obtain from the variation $\delta S_{\phi}/\delta\phi=0$ as
$$
\Box\phi-\xi R\phi-\ve \,U'(\phi)=0\,
$$
where $()'=\frac{\ud}{\ud\phi}$.

Assuming the spatially flat Friedmann-Robertson-Walker metric
$$
\ud s^{2}=-\ud t^{2}+a^{2}(t)\Big(\ud x^{2}+\ud y^{2}+\ud z^{2}\Big)\,.
$$
we obtain the following energy conservation condition
\begin{equation}
\label{eq:constr}
\frac{3}{\kappa^{2}}H^{2}=\rho_{\phi}=\ve\frac{1}{2}\dot{\phi}^{2}+U(\phi)+\ve3\xi H^{2}\phi^{2}+\ve6\xi H \phi\dot{\phi}\,,
\end{equation}
the acceleration equation
\begin{equation}
\label{eq:accel}
\dot{H}=-2H^{2}+\frac{\kappa^{2}}{6}\frac{-\ve(1-6\xi)\dot{\phi}^{2}+4U(\phi)-6\xi\phi U'(\phi)}{1-\ve\xi(1-6\xi)\kappa^{2}\phi^{2}}\,,
\end{equation}
and the equation of motion for the scalar field
\begin{equation}
\label{eq:sc_field}
\ddot{\phi}+3H\dot{\phi} + 6\xi\big(\dot{H}+2H^{2}\big)\phi + \ve U'(\phi)=0\,,
\end{equation}
where an over dot denotes differentiation with respect to the cosmological time.

Dynamical equations \eqref{eq:accel} and \eqref{eq:sc_field} subject to the energy conservation condition \eqref{eq:constr} with a given scalar field potential function completely describe dynamical evolution of the model. One can introduce various dimensionless dynamical variables \cite{Hrycyna:2010yv,Hrycyna:2015eta,Hrycyna:2020jmw,Kerachian:2019tar,Jarv:2021qpp} in order to parametrise the phase space and investigate dynamics of a model. In the previous paper \cite{Hrycyna:2020jmw} it was shown that for a monomial potential functions at infinite values of the scalar field there exist generic de Sitter and Einstein--de Sitter states for a particular value of the non-minimal coupling constant $\xi=\frac{3}{16}$. In the present paper we are interested only in asymptotic behaviour of a limited class of scalar field potential function, namely, a class with asymptotic in the form of a quadratic potential function \cite{Hrycyna:2007mq, Hrycyna:2008gk, Jarv:2021qpp}.

\section{An asymptotically quadratic potential function}

In this section we investigate dynamics with an asymptotically quadratic scalar field potential $U(\phi)\to\pm\frac{1}{2}m^{2}\phi^{2}$ as $\phi\to\infty$.

Let us assume, that starting from some values of the scalar field $\phi>\phi^{*}$ the potential function can be approximated as
$$
U(\phi)=\pm\frac{1}{2}m^{2}\phi^{2} \pm M^{4+n}\phi^{-n}\,,
$$
where $n>-2$ and the second term constitutes small, asymptotically vanishing deviation. The scalar field potential function under investigations can be treated as the assumed approximation to potential energy at large $\phi$ and also as a classical phenomenological modification and extension to the Ratra–Peebles potential \cite{Peebles:1987ek, Ratra:1987rm}.

Next, we introduce the following dimensionless phase space variables
$$
u=\frac{\dot{\phi}}{H\phi}\,,\quad v=\frac{\sqrt{6}}{\kappa}\frac{1}{\phi}\,,
$$
and dimensionless parameters describing potential function of the scalar field
$$
\mu=\pm\frac{m^{2}}{H_{0}^{2}}\,,\quad \alpha=\pm2\frac{M^{4+n}}{H_{0}^{2}}\left(\frac{\kappa}{\sqrt{6}}\right)^{2+n}\,,
$$
where $H_{0}$ is the present value of the Hubble function.

Thus we obtain the energy conservation condition \eqref{eq:constr} in the form
\begin{equation}
\label{eq:en_cons}
\frac{H^{2}}{H_{0}^{2}}=\frac{\mu+\alpha v^{2+n}}{v^{2}-\ve(1-6\xi)u^{2}-\ve6\xi(u+1)^{2}}
\end{equation}
and the acceleration equation \eqref{eq:accel} as
\begin{equation}
\label{eq:accel_dS}
\begin{split}
\frac{\dot{H}}{H^{2}}=&-2+\frac{1}{v^{2}-\ve6\xi(1-6\xi)}\Bigg(-\ve(1-6\xi)u^{2} \\ &+\left(\frac{H^{2}}{H_{0}^{2}}\right)^{-1}\left(2(1-3\xi)\mu+(2+3\xi n)\alpha v^{2+n}\right)\Bigg)\,.
\end{split}
\end{equation}
Finally, the dynamical system describing evolution of the model under consideration is in the following form
\begin{equation}
\label{eq:dyn_sys}
\begin{split}
\frac{\ud u}{\ud \ln{a}} = &-u(u+1)-(u+6\xi)\left(\frac{\dot{H}}{H^{2}}+2\right)\\ &  -\ve\left(\frac{H^{2}}{H_{0}^{2}}\right)^{-1}\left(\mu-\frac{1}{2}n \alpha v^{2+n}\right)\,,\\
\frac{\ud v}{\ud \ln{a}} = &-uv\,,
\end{split}
\end{equation}
where evolution of the phase space curves is a function of the natural logarithm of the scale factor.

In what follows, we are interested only in asymptotic states located at infinite values of the scalar field $\phi$ i.e. with $v^{*}\equiv0$. One can easily check that for $\xi\ne\left\{0,\frac{1}{6},\frac{1}{4}\right\}$ there are three such states with $u^{*}=-6\xi\pm\sqrt{-6\xi(1-6\xi)}$ and $u^{*}=-\frac{2\xi}{1-4\xi}$, with the later one giving rise to de Sitter type of evolution.

\section{Instability of the initial de Sitter state}

We will investigate dynamical behaviour in the vicinity of the critical point
$$
u^{*}= - \frac{2\xi}{1-4\xi}\,, \quad v^{*}=0\,.
$$ 
The energy conservation condition \eqref{eq:en_cons} calculated at this point gives
\begin{equation}
\label{eq:en_dS}
\frac{H^{2}}{H_{0}^{2}}\bigg|^{*}=-\ve\mu\frac{(1-4\xi)^{2}}{2\xi(1-6\xi)(3-16\xi)}>0\,,
\end{equation}
and this quantity must be positive in order to obtain the asymptotic state in the physical region of the phase space. With the vanishing acceleration equation \eqref{eq:accel_dS} at this critical point and the constant value of the energy conservation condition \eqref{eq:en_dS} we conclude that this state corresponds to the de Sitter type of evolution. 

To determine stability conditions of the critical point under considerations we calculate eigenvalues of linearisation matrix of the system \eqref{eq:dyn_sys} which gives
$$
\lambda_{1}=-4+\frac{1}{1-4\xi}\,,\quad \lambda_{2}=\frac{2\xi}{1-4\xi}\,,
$$
and the linearised solutions in the following form
\begin{equation}
\begin{split}
u(a) & = u^{*} +\Delta u \left(\frac{a}{a^{(i)}}\right)^{\lambda_{1}}\,,\\
v(a) & = \Delta v \left(\frac{a}{a^{(i)}}\right)^{\lambda_{2}}\,,
\end{split}\nonumber
\end{equation}
where $\Delta u = u^{(i)}-u^{*}$ and $\Delta v= v^{(i)}$ are the initial conditions in the vicinity of the state.

The stability conditions will be established with respect to expansion of the model since the time parameter in dynamical equations \eqref{eq:dyn_sys} is natural logarithm of the scale factor. Combining the eigenvalues of the linearisation matrix with positivity of the energy conservation condition at the critical point \eqref{eq:en_dS} we obtain the following stability conditions of the de Sitter state:
\begin{itemize}
\item[--]{a stable node with $\lambda_{1}<0\, \wedge\, \lambda_{2}<0$ for
\begin{equation}
\bigg(\ve\mu<0\, \wedge\, \xi>\frac{1}{4}\bigg)\, \lor\, 
\bigg(\ve\mu>0\, \wedge\, \xi<0\bigg)\,, \nonumber
\end{equation}}
\item[--]{a saddle with $\lambda_{1}<0\, \wedge\, \lambda_{2}>0$ for 
\begin{equation}
\bigg(\ve\mu<0\, \wedge\, 0<\xi<\frac{1}{6}\bigg)\, \lor\, 
\bigg(\ve\mu>0\, \wedge\, \frac{1}{6}<\xi<\frac{3}{16}\bigg)\,,\nonumber
\end{equation}}
\item[--]{an unstable node with $\lambda_{1}>0\, \wedge\, \lambda_{2}>0$ for 
\begin{equation}
\ve\mu<0\, \wedge\, \frac{3}{16}<\xi<\frac{1}{4}\,.
\nonumber
\end{equation}}
\end{itemize}
We conclude that unstable de Sitter state exists for the non-minimal coupling constant $\frac{3}{16}<\xi<\frac{1}{4}$ both for the canonical and the phantom scalar field. Additionally we obtain the asymptotic value of the scalar field potential function from the condition $\ve\mu<0$ and for the canonical scalar field with $\ve=+1$ potential tends to negative values $U(\phi)\to -\frac{1}{2}m^{2}\phi^{2}$ \cite{Felder:2002jk,Boisseau:2015hqa,Hrycyna:2020jmw} while for the phantom scalar field with $\ve=-1$ potential tends to positive values $U(\phi)\to +\frac{1}{2}m^{2}\phi^{2}$.

We should notice some analogies between asymptotic state under consideration and fast-roll (or rapid-roll) inflationary state \cite{Linde:2001ae,Kofman:2007tr,Chiba:2008ia}. The condition for rapid-roll inflation with the conformal coupling $\xi=\frac{1}{6}$ is $\dot{\phi}=-H\phi$ \cite{Kofman:2007tr} which corresponds to the dynamical variable $u=-1$. In our case this value of the non-minimal coupling is excluded since it leads to a non-hyperbolic, degenerated asymptotic state. For the de Sitter state under considerations the fast-roll condition is $u=-\frac{2\xi}{1-4\xi}$ which gives the behaviour of the scalar field as $\dot{\phi}=-\frac{2\xi}{1-4\xi}H\phi$ and we obtain that for $\frac{3}{16}<\xi<\frac{1}{4}$ at infinite values of the scalar field $\phi\to+\infty$ the scalar field changes infinitely fast with respect to the cosmological time $\dot{\phi}\to-\infty$.

Also, we should be aware that if the de Sitter state is reached asymptotically, then the model might not be non-singular automatically. The scalar curvature invariants would be finite in this case, but a parallelly propagated singularity could be present \cite{Yoshida:2018ndv,Nomura:2021lzz}. Using the linearised solutions to dynamics in vicinity of the de Sitter state we find
$$
\frac{\dot{H}}{a^{2}}\left(\frac{a^{(i)}}{H_{0}}\right)^{2} \approx 2 \frac{H^{2}}{H_{0}^{2}}\bigg|^{*} \Delta u \left(\frac{a}{a^{(i)}}\right)^{-6+\frac{1}{1-4\xi}}\,,
$$
and at the de Sitter state represented by an unstable node in the limit $a\to0$ this quantity diverges to infinity for $\frac{3}{16}<\xi<\frac{5}{24}$, tends to a constant value for $\xi=\frac{5}{24}$, and vanishes for $\frac{5}{24}<\xi<\frac{1}{4}$. This way we have found more stringent constraint on the non-minimal coupling constant in the model giving rise to non-singular beginning of the universe.

\section{Physics from dynamics}

The general form of the energy conservation condition in a model with the FRW symmetry and an arbitrary form of matter can be presented as
$$
\frac{3}{\kappa^{2}}H^{2}=\rho_{\text{eff}}<\rho_{\text{Pl}}=m_{\text{Pl}}^{4}\,,
$$
where we need to assume that during the evolution the effective energy density 
$\rho_{\text{eff}}$ is smaller than the Planck energy density in order to eliminate possible contributions form quantum gravity effects.
 
From the energy conservation condition \eqref{eq:en_dS} we have that at the asymptotic unstable de Sitter state
$$
\frac{H^{2}}{H_{0}^{2}}\bigg|^{*}=-\ve\mu\frac{(1-4\xi)^{2}}{2\xi(1-6\xi)(3-16\xi)}< \frac{\kappa^{2}m_{\text{Pl}}^{4}}{3H_{0}^{2}}\,,
$$
where $\ve\mu<0$ and we obtain the following inequality 
$$
\frac{m^{2}}{m_{\text{Pl}}^{2}}\frac{3}{8\pi}\frac{(1-4\xi)^{2}}{2\xi(1-6\xi)(3-16\xi)}<1\,.
$$
We can observe that even for the mass of the scalar field of order of the Planck mass $m^{2}\simeq m_{\text{Pl}}^{2}$ we have
$$
\frac{3}{8\pi}\frac{(1-4\xi)^{2}}{2\xi(1-6\xi)(3-16\xi)}<1\,,
$$
and we can find values of the non-minimal coupling constant that satisfy this relation and potential quantum gravity effects are excluded from the model.

\section{The Einstein frame dynamical analysis}

To obtain some considerable mathematical simplification of the theory one usually uses the following conformal transformation of the metric tensor 
$$
\tilde{g}_{\mu\nu}=\Omega^{2}g_{\mu\nu}\,,
$$
with the conformal factor
\begin{equation}
\label{eq:confac}
\Omega^{2}=\big|1-\ve\xi\kappa^{2}\phi^{2}\big|\,,
\end{equation}
where $\big|.\big|$ stands for the absolute value of an argument, in order to relate cosmological models with a non-minimally coupled scalar field in, so called, the Jordan frame with its conformal counterpart with, now, a minimally coupled scalar field in so called the Einstein frame. Note that the values of the scalar field $\phi$ where the conformal factor vanishes are excluded. 

Now, the total action integral for the theory in the Jordan frame \eqref{eq:action} is
\begin{equation}
S = \sgn\left(1-\ve\xi\kappa^{2}\phi^{2}\right)\left(\tilde{S}_{\tilde{g}}+\tilde{S}_{\phi}\right)\,,
\end{equation}
where $\sgn(.)$ stands for the sign function and action integrals for the theory in the Jordan frame \eqref{eq:j_grav} and \eqref{eq:j_m} transform to
\begin{equation}
\label{eq:e_grav}
\tilde{S}_{\tilde{g}} = \frac{1}{2\kappa^{2}}\int\ud^{4}x\sqrt{-\tilde{g}}\tilde{R}\,,
\end{equation}
and
\begin{equation}
\label{eq:e_m}
\tilde{S}_{\phi} = -\frac{1}{2}\int\ud^{4}x\sqrt{-\tilde{g}}\left(\frac{\omega(\phi)}{\phi}\tilde{\nabla}^{\alpha}\phi\,\tilde{\nabla}_{\alpha}\phi + 2\tilde{U}(\phi)\right)\,,
\end{equation}
where
$$
\omega(\phi) = \ve\frac{1-\ve\xi(1-6\xi)\kappa^{2}\phi^{2}}{\big(1-\ve\xi\kappa^{2}\phi^{2}\big)^{2}}\phi\,,
$$
and the scalar field potential function
$$
\tilde{U}(\phi)= \sgn\left(1-\ve\xi\kappa^{2}\phi^{2}\right)\frac{U(\phi)}{\big(1-\ve\xi\kappa^{2}\phi^{2}\big)^{2}}\,.
$$
Using the following scalar field transformation 
$$
\ud\tilde{\phi} = \frac{\sqrt{\ve\big(1-\ve\xi(1-6\xi)\kappa^{2}\phi^{2}\big)}}{1-\ve\xi\kappa^{2}\phi^{2}}\ud\phi\,
$$
one can normalise scalar field kinetic term in \eqref{eq:e_m} and work in canonical minimally coupled scalar field cosmology. In the present approach we work in original scalar field dynamical variable in order to directly compare dynamical behaviour of the model in Jordan and Einstein frames.

We have to note that the minimally coupled phantom scalar field with $\ve=-1$ give rise to nonphysical ghosts (see e.g. \cite{Cline:2003gs}). However, in theory under considerations with non-minimally coupled scalar field given by the Einstein frame action integrals \eqref{eq:e_grav} and \eqref{eq:e_m} we have the following condition for the theory to be ghost free
$$
\ve\big(1-\ve\xi(1-6\xi)\kappa^{2}\phi^{2}\big)>0\,.
$$
Thus, both for the canonical $\ve=+1$ and the phantom $\ve=-1$ non-minimally coupled scalar field we can always find values of the non-minimal coupling constant $\xi$ and phase space regions where the theory is ghost free.

In the Einstein frame the field equations are the following
$$
\tilde{R}_{\mu\nu}-\frac{1}{2}\tilde{g}_{\mu\nu}\tilde{R}=\kappa^{2}\tilde{T}^{(\phi)}_{\mu\nu}\,,
$$
where now the energy momentum tensor for the scalar field is
$$
\tilde{T}^{(\phi)}_{\mu\nu}=\frac{\omega(\phi)}{\phi}\tilde{\nabla}_{\mu}\phi\,\tilde{\nabla}_{\nu}\phi -\frac{1}{2}\frac{\omega(\phi)}{\phi}\tilde{g}_{\mu\nu}\tilde{\nabla}^{\alpha}\phi \, \tilde{\nabla}_{\alpha}\phi - \tilde{U}(\phi)\tilde{g}_{\mu\nu}\,,
$$
and from variation $\delta\tilde{S}_{\phi}/\delta\phi=0$ we have
$$
2\frac{\omega(\phi)}{\phi}\tilde{\Box}\phi +\left(\frac{\omega'(\phi)}{\phi}-\frac{\omega(\phi)}{\phi^{2}}\right)\tilde{\nabla}^{\alpha}\phi\,\tilde{\nabla}_{\alpha}\phi-2\tilde{U}'(\phi)=0\,,
$$

Now, we can assume the Einstein frame spatially flat FRW metric
$$
\ud\tilde{s}^{2} = -\ud\tilde{t}^{2} + \tilde{a}^{2}\left(\tilde{t}\right)\Big(\ud x^{2}+\ud y^{2}+\ud z^{2}\Big)\,,
$$
where the cosmological time and the scale factor transform according to
$$
\ud\tilde{t} = \Omega \ud t\,, \qquad \tilde{a} = \Omega a\,.
$$
We introduce the following dimensionless phase space variables
$$
\tilde{u}=\frac{\dot{\phi}}{\tilde{H}\phi}\,,\quad v=\frac{\sqrt{6}}{\kappa}\frac{1}{\phi}\,,
$$
where now a dot denotes differentiation with respect to time $\tilde{t}$ and $\tilde{H}$ is the Einstein frame Hubble function given by 
$$
\tilde{H} = \frac{1}{\tilde{a}(\tilde{t})}\frac{\ud\tilde{a}(\tilde{t})}{\ud\tilde{t}}\,.
$$
Following the Jordan frame approach the dimensionless parameters describing potential function of the scalar field are introduced
$$
\tilde{\mu}=\pm\frac{m^{2}}{\tilde{H}_{0}^{2}}\,,\quad \tilde{\alpha}=\pm2\frac{M^{4+n}}{\tilde{H}_{0}^{2}}\left(\frac{\kappa}{\sqrt{6}}\right)^{2+n}\,,
$$
where $\tilde{H}_{0}$ is the present value of the Hubble function in the Einstein frame. Thus, we obtain the Einstein frame energy conservation condition
\begin{equation}
\label{eq:ef_energy}
\frac{\tilde{H}^{2}}{\tilde{H}_{0}^{2}} = 
\frac{\sgn\left(v^{2}-\ve6\xi\right)\,v^{2}\big(\tilde{\mu} + \tilde{\alpha} v^{n+2}\big)}{\big(v^{2}-\ve6\xi\big)^{2}-\ve\big(v^{2}-\ve6\xi(1-6\xi)\big)\tilde{u}^{2}}\,,
\end{equation}
and the acceleration equation
\begin{equation}
\label{eq:ef_accel}
\frac{\dot{\tilde{H}}}{\tilde{H}^{2}} = -\ve3\frac{v^{2}-\ve6\xi(1-6\xi)}{\big(v^{2}-\ve6\xi\big)^{2}}\tilde{u}^{2}\,.
\end{equation}
Dynamical system describing evolution of the model can be directly found using the procedure from the Jordan frame approach, but now, the evolution of the phase space curves is in the Einstein frame scale factor $\tilde{a}$. Again, we are interested only in the critical points located at infinite values of the scalar field, i.e. at $v^{*}\equiv0$. We find three such critical points, for $\xi\ne\frac{1}{6}$, with $\tilde{u}^{*}=\pm\sqrt{-\frac{6\xi}{1-6\xi}}$ and $\tilde{u}^{*}=-\frac{2\xi}{1-6\xi}$ with the later one giving rise to the Einstein static universe solution. 

The asymptotic state under considerations 
$$
\tilde{u}^{*}= - \frac{2\xi}{1-6\xi}\,, \quad v^{*}=0\,,
$$
gives the vanishing energy conservation condition \eqref{eq:ef_energy} and the acceleration equation \eqref{eq:ef_accel} as
$$
\frac{\dot{\tilde{H}}}{\tilde{H}^{2}}\bigg|^{*} = \frac{2\xi}{1-6\xi}\,.
$$
One can conclude that this leads to vanishing of the Einstein frame cosmological time derivative of the Hubble function 
$$
\frac{\dot{\tilde{H}}}{\tilde{H}^{2}_{0}}\bigg|^{*} = 0\,,
$$
and give rise to the Einstein static universe.

The eigenvalues of the linearisation matrix at this state are
$$
\lambda_{1}=-\frac{3-16\xi}{1-6\xi}\,, \quad \lambda_{2}=\frac{2\xi}{1-6\xi}\,,
$$
and then the linearised solutions are
\begin{equation}
\begin{split}
\tilde{u}(\tilde{a}) & = \tilde{u}^{*} +\Delta\tilde{u} \left(\frac{\tilde{a}}{\tilde{a}^{(i)}}\right)^{\lambda_{1}}\,,\\
v(\tilde{a}) & = \Delta v \left(\frac{\tilde{a}}{\tilde{a}^{(i)}}\right)^{\lambda_{2}}\,,
\end{split}\nonumber
\end{equation}
where $\Delta\tilde{u} = \tilde{u}^{(i)}-\tilde{u}^{*}$ and $\Delta v= v^{(i)}$ are the initial conditions in the vicinity of the state and $\tilde{a}^{(i)}$ is the initial value of the scale factor in the Einstein frame.

\begin{figure*}
\centering
\includegraphics[scale=0.5]{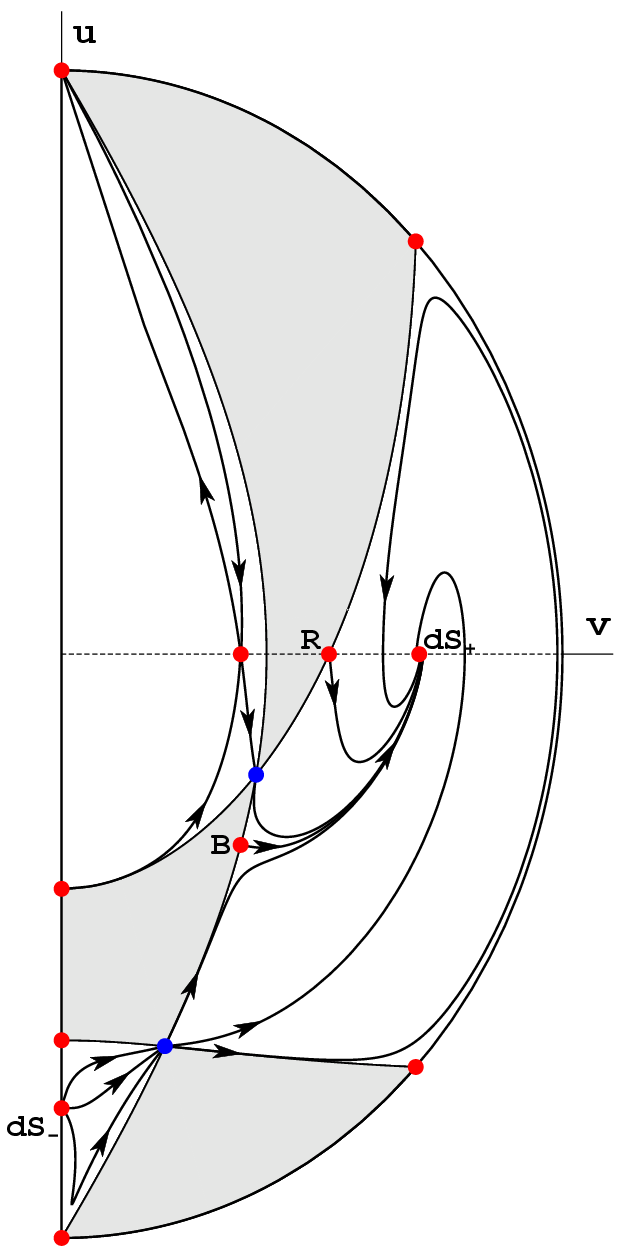}
\hspace{3cm}
\includegraphics[scale=0.5]{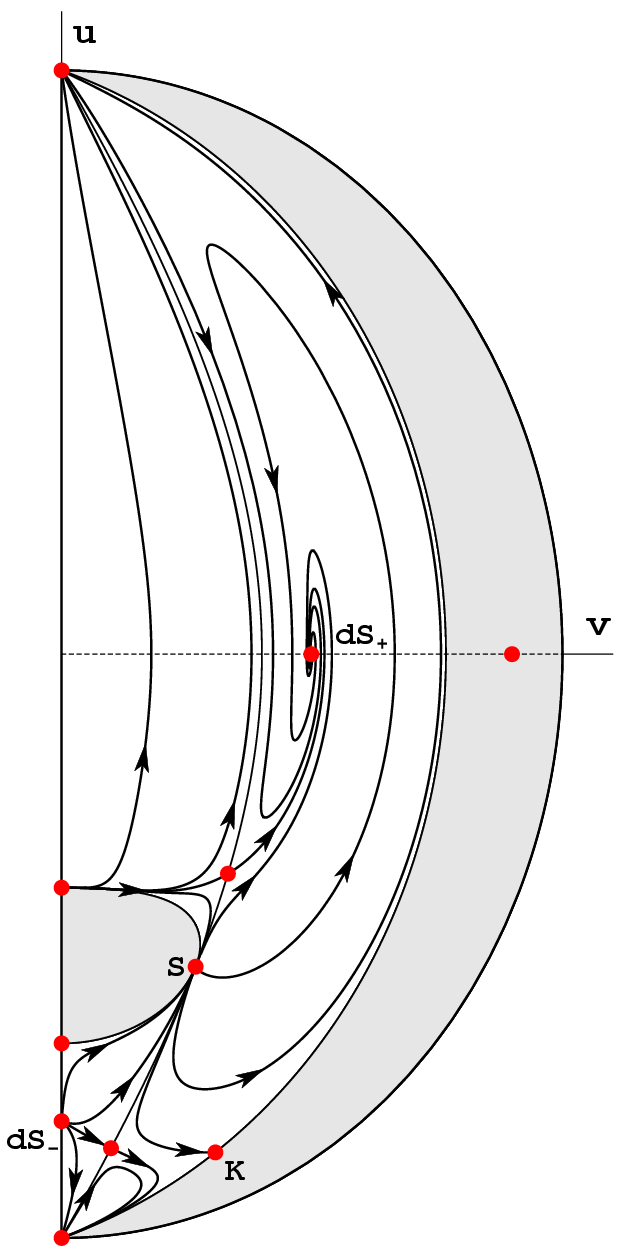}\\
\includegraphics[scale=0.5]{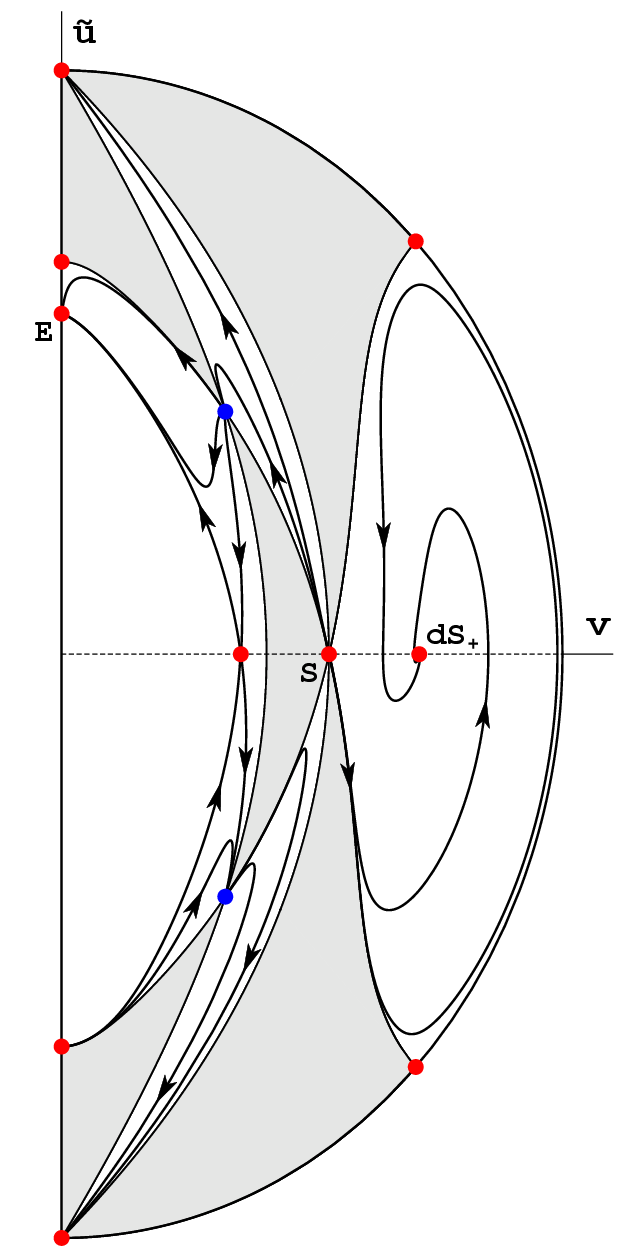}
\hspace{3cm}
\includegraphics[scale=0.5]{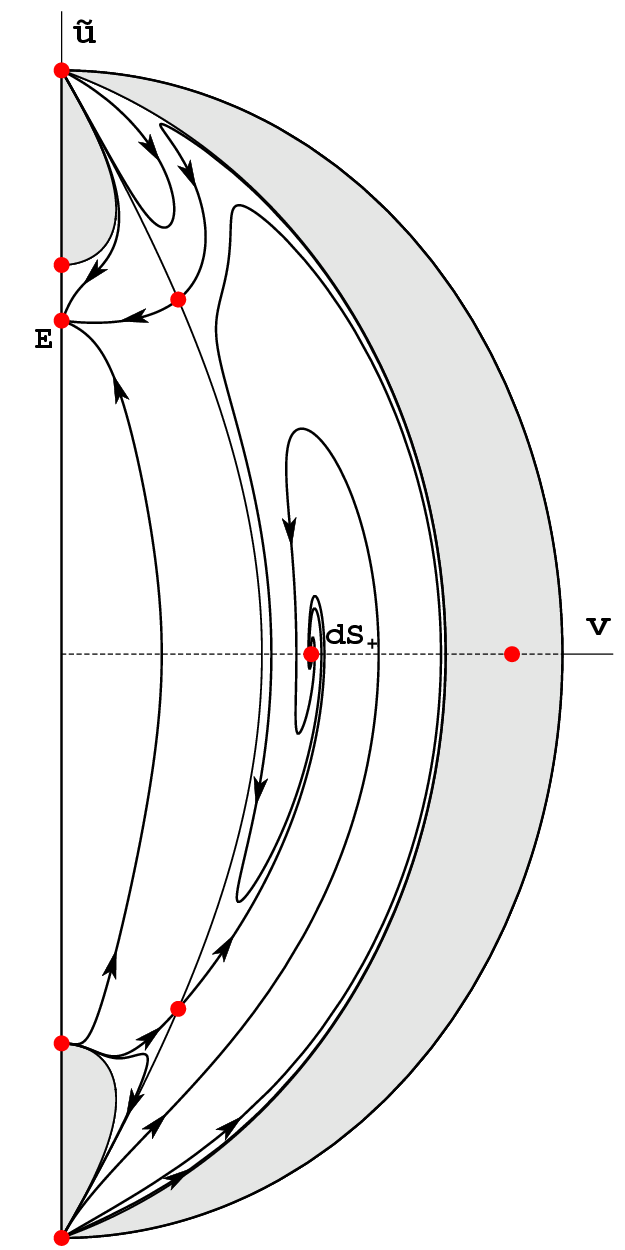}
\caption{The phase space diagrams representing evolutional paths of the dynamical system \eqref{eq:dyn_sys} compactified with circle at infinity in the Jordan frame (top) and  in the Einstein frame (bottom) with $\ve=+1$, $\xi=\frac{7}{32}$, $n=1$, $\mu=-1$, $\alpha=3$ (left) and $\ve=-1$, $\xi=\frac{2}{9}$, $n=-\frac{1}{2}$, $\mu=1$, $\alpha=-\frac{1}{6}$ (right). The direction of arrows on the phase space trajectories indicate expansion of universe. The vertical line through critical point $S$ on the bottom left phase space diagram corresponds to vanishing of the conformal factor for the values of the scalar field $\kappa^{2}\phi^{2}=\frac{1}{\ve\xi}$. The shaded regions where the Hubble functions $H^{2}<0$ or $\tilde{H}^{2}<0$ are nonphysical. We immediately see that both formulation of the model, for the canonical and the phantom scalar field, are physically nonequivalent since the unstable de Sitter state in the Jordan frame $dS_{-}$ is transformed in the stable Einstein universe in the Einstein frame $E$.}
\label{fig:1}
\end{figure*}

The stability conditions of the asymptotic Einstein static universe are obtained form the eigenvalues of the linearisation matrix calculated at this state and we find that the critical point is in the form of
\begin{itemize}
\item[--]{a stable node with $\lambda_{1}<0\, \wedge\, \lambda_{2}<0$ for 
$$\xi<0\, \lor\, \xi>\frac{3}{16}\,,$$}
\item[--]{a saddle with $\lambda_{1}<0\, \wedge\, \lambda_{2}>0$\, $\lor$\, $\lambda_{1}>0\, \wedge\, \lambda_{2}<0$ for 
$$0<\xi<\frac{3}{16}\,.$$}
\end{itemize}
Note that a minimally coupled scalar field case with $\xi=0$ is excluded from the considerations since for this value both frames of the theory coincide. Additionally, for $\xi=\frac{3}{16}$ which is the value of conformally coupled scalar field in $5$D theory of gravity \cite{Hrycyna:2020jmw}, the first eigenvalue $\lambda_{1}$ vanishes what leads to a degenerated non-hyperbolic critical point.

Additionally, using the condition for the Einstein frame theory to be ghost free we find that for values of the non-minimal coupling constant under considerations $\frac{3}{16}<\xi<\frac{1}{4}$ the asymptotically stable Einstein static solution is located in the ghost free region of the phase space.

Now, we can show that the asymptotic Einstein static universe in the Einstein frame indeed corresponds to the de Sitter asymptotic state in the Jordan frame. The phase space variables in the Einstein and the Jordan frame are
$$
\tilde{u}= \frac{\frac{\ud\phi}{\ud\tilde{t}}}{\tilde{H}\phi}\,,\qquad u = \frac{\frac{\ud\phi}{\ud t}}{H\phi}\,,
$$
using the following transformations between the frames
$$
\ud\tilde{t}=\Omega\ud t\,, \qquad \tilde{H}=\frac{1}{\Omega}\bigg(H+\frac{1}{\Omega}\frac{\ud\Omega}{\ud t}\bigg)\,,
$$
we obtain 
$$
\frac{u}{\tilde{u}} = 1+\frac{1}{2\Omega^{2}}\frac{\ud\Omega^{2}}{\ud\ln{a}}\,.
$$
The conformal factor in the theory is given by \eqref{eq:confac} and we find the following equation relating phase space variables in the Jordan frame and the Einstein frame
$$
\frac{u}{\tilde{u}} = 1 - \frac{\ve6\xi}{v^{2}-\ve6\xi}u\,.
$$
At the asymptotic states under considerations $v^{*}\equiv0$ this relation reduces to the following transformations between both frames 
\begin{equation}
\begin{split}
u^{*} & =  \frac{\tilde{u}^{*}}{1-\tilde{u}^{*}}\,,\\
\tilde{u}^{*} &=  \frac{u^{*}}{1+u^{*}}\,.
\end{split}\nonumber
\end{equation}
Finally, we arrive at the following correspondence between asymptotic states in both frames
$$
u^{*}= - \frac{2\xi}{1-4\xi} \quad \iff \quad \tilde{u}^{*}= - \frac{2\xi}{1-6\xi}\,,
$$
and we can conclude that for the non-minimal coupling constant $\frac{3}{16}<\xi<\frac{1}{4}$ the unstable de Sitter state in the Jordan frame corresponds to the stable Einstein static state in the Einstein frame.

Figure~\ref{fig:1} represents the phase space diagrams of the dynamical system \eqref{eq:dyn_sys} compactified with circle at infinity in the Jordan frame and in the Einstein frame, both, for the canonical $\ve=+1$ and the phantom $\ve=-1$ scalar field. The direction of arrows on the phase space trajectories indicate expansion of universe. The blue dots in the phase space diagrams for the canonical scalar field (left panel) do not correspond to any particular physical state since the evolution at those states is regular with finite values of the Hubble function and the acceleration equation, additionally, the scalar field potential function vanishes at those points. One can immediately notice differences in evolution in the vicinity of the asymptotic states under considerations, namely, the unstable de Sitter states $dS_{-}$ in the Jordan frame and the stable Einstein static solutions $E$ in the Einstein frame. We conclude that both formulations of the model are physically nonequivalent since the unstable de Sitter state in the Jordan frame is transformed in the stable Einstein universe in the Einstein frame. 

\section{The special case $\xi=\frac{1}{4}$}

In this section we investigate dynamics for the special, previously excluded value of the non-minimal coupling constant $\xi=\frac{1}{4}$ for which location of the investigated asymptotic state is moved to infinity. Examination of the energy conservation condition  \eqref{eq:en_dS} indicates that for this value of the parameter the physics of the asymptotic state can be completely different from the de Sitter state.

Introducing the following projective coordinate
$$\hat{u}=\frac{1}{u}=H\frac{\phi}{\dot{\phi}}$$
in the dynamical system \eqref{eq:dyn_sys} and the following time redefinition
\begin{equation}
\frac{\ud}{\ud\eta}=-\ve\hat{u}\left(v^{2}+\ve\frac{3}{4}\right)\left(1+\frac{\alpha}{\mu}v^{2+n}\right)\frac{\ud}{\ud\ln{a}}\,,
\label{eq:time_spec}
\end{equation}
with the energy conservation condition
$$
\frac{H^{2}}{H_{0}^{2}}=\hat{u}^{2}\frac{\mu+\alpha v^{2+n}}{v^{2}\hat{u}^{2}+\ve\frac{1}{2}-\ve\frac{3}{2}(1+\hat{u})^{2}}\,,
$$
and the acceleration equation
$$
\frac{\dot{H}}{H^{2}} = -2+ \frac{\ve\frac{1}{2}+\hat{u}^{2}\left(\frac{H^{2}}{H_{0}^{2}}\right)^{-1}\left(\frac{1}{2}\mu+\left(2+\frac{3}{4}n\right)\alpha v^{2+n}\right)}{\hat{u}^{2}\left(v^{2}+\ve\frac{3}{4}\right)}\,,
$$
we find dynamical system with the new projective coordinates.

In what follows we are interested in dynamics in vicinity of the critical point 
$$
\hat{u}^{*}= 0\,, \quad v^{*}=0\,,
$$
with linearised solutions
\begin{equation}
\begin{split}
\hat{u}(\eta) & = \Delta \hat{u} \exp{\left(\frac{3}{2}\eta\right)}\,,\\
v(\eta) & = \Delta v \exp{\left(\frac{3}{4}\eta\right)}\,,
\end{split}
\label{eq:sol_spec}
\end{equation}
where $\Delta\hat{u} = \hat{u}^{(i)}$ and $\Delta v= v^{(i)}$ are the initial conditions. The critical point is unstable in time $\eta$. 

Simple inspection gives that at the asymptotic state the energy conservation condition vanishes
together with the cosmological time derivative of the Hubble function
\begin{equation}
\frac{H^{2}}{H_{0}^{2}}\bigg|^{*} = 0\,, \quad \frac{\dot{H}}{H_{0}^{2}}\bigg|^{*} = 0\,,
\end{equation}
which gives rise to the Einstein static universe \cite{Ellis:2002we,Ellis:2003qz,Barrow:2003ni}. Additionally, the acceleration equation calculated at this state is
\begin{equation}
\frac{\dot{H}}{H^{2}}\bigg|^{*} = -3\,,
\end{equation}
which suggest that the Einstein static state under considerations is filled with effective substance in the form of Zeldovich stiff matter with equation of state parameter $w_{\text{eff}}=1$ \cite{Zeldovich:1962,Zeldovich:1972zz}.

\begin{figure*}
\centering
\includegraphics[scale=0.5]{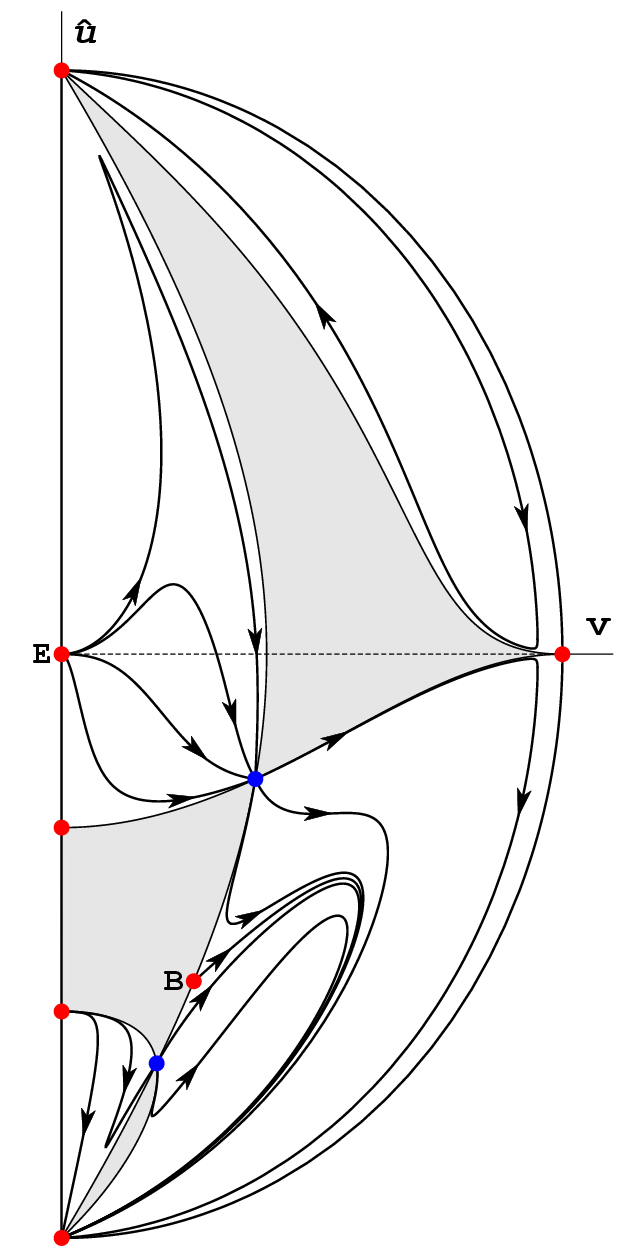}
\hspace{3cm}
\includegraphics[scale=0.5]{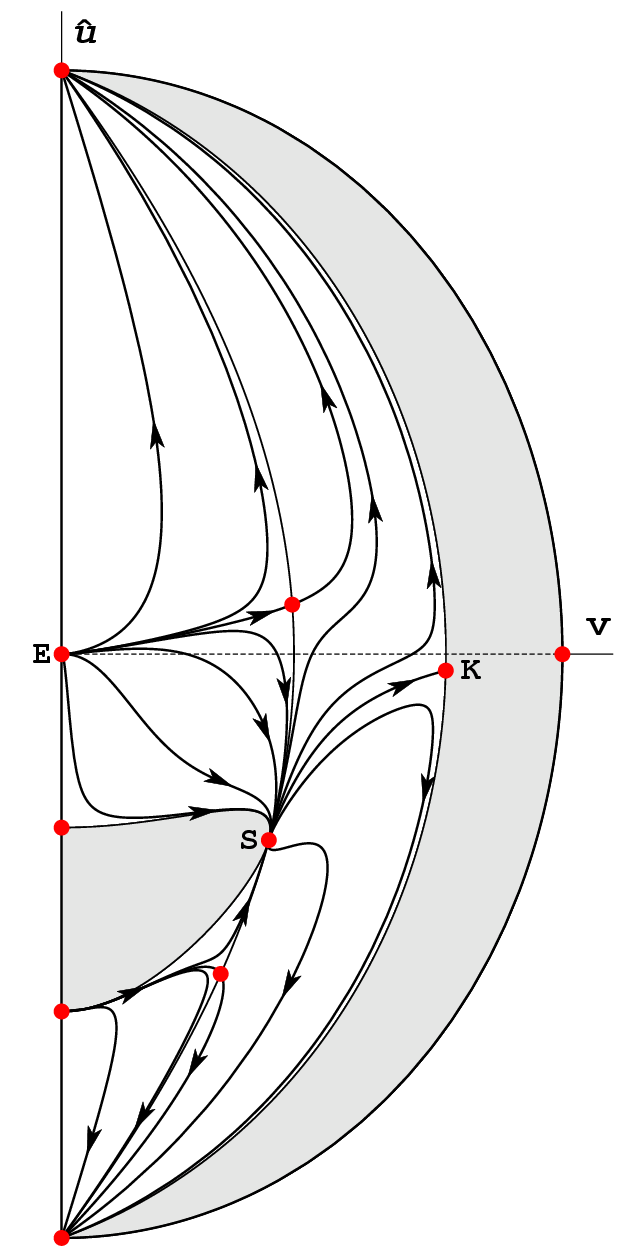}\\
\includegraphics[scale=0.5]{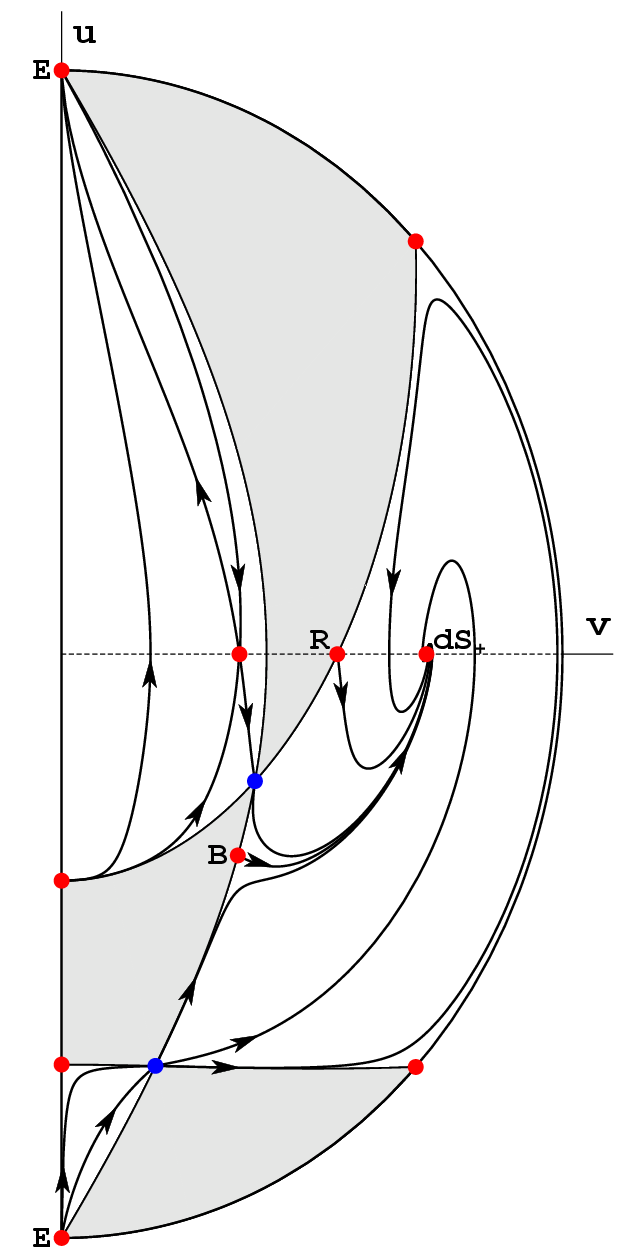}
\hspace{3cm}
\includegraphics[scale=0.5]{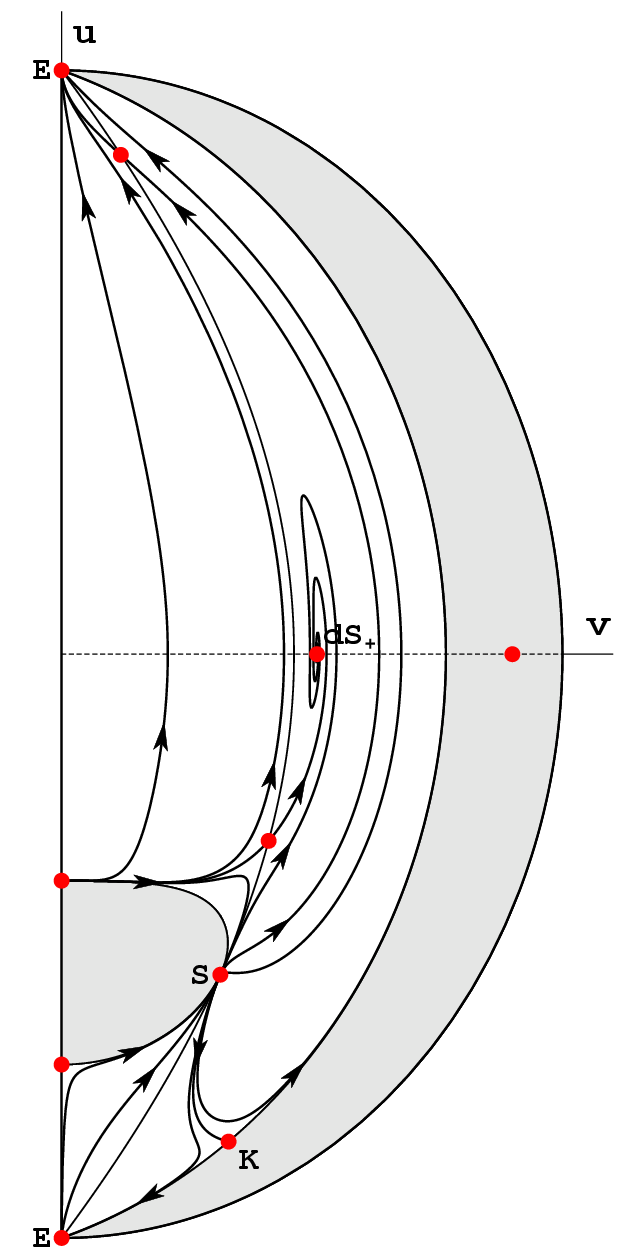}
\caption{The phase space diagrams representing evolutional paths of the dynamical system \eqref{eq:dyn_sys} for the special case with $\xi=\frac{1}{4}$ in the $(v\,,\hat{u})$ variables where direction of the arrows corresponds to time $\eta$ and the original $(v\,,u)$ dynamical variables where direction of arrows corresponds to the scale factor $a$.  The phase space diagrams are compactified with circles at infinity of phase spaces. The left panel represents the phase space diagrams for the canonical scalar field $\ve=+1$ with $n=1$, $\mu=-1$, $\alpha=3$  while the right panel for the phantom scalar field $\ve=-1$, with $n=-\frac{1}{2}$, $\mu=1$, $\alpha=-\frac{1}{6}$. The critical point $E$ corresponds to the Einstein static universe while the critical point $dS_{+}$ denotes the stable de Sitter state. From the phase space diagrams in the left panel we notice open and dense set of initial conditions leading to the non-singular evolutional paths connecting the unstable Einstein static universe $E$ with the stable de Sitter state $dS_{+}$. The shaded regions where $H^{2}<0$ are nonphysical.}
\label{fig:2}
\end{figure*}

Up to linear terms the time redefinition \eqref{eq:time_spec} is in the following form
$$
\ud\ln{a} \approx -\frac{3}{4}\hat{u}(\eta)\ud\eta\,,
$$
using the linearised solutions \eqref{eq:sol_spec} we obtain
$$
\int_{a^{(i)}}^{a^{*}}\frac{1}{a}\ud a \approx -\frac{3}{4}\Delta\hat{u}\int_{0}^{-\infty}\exp{\left(\frac{3}{2}\eta\right)}\ud\eta\,,
$$
where integration is taken from the initial state toward the critical point since the critical point under considerations in the form of an unstable node. Thus, we obtain
$$
\ln{\left(\frac{a^{*}}{a^{(i)}}\right)}\approx\frac{1}{2}\Delta\hat{u}\,,
$$
and one can read that value of the scale factor at the critical point corresponding to the Einstein static universe is
\begin{equation}
\begin{split}
a^{*}& > a^{(i)} \quad \text{for} \quad \Delta\hat{u}>0\,,\\
a^{*}& < a^{(i)} \quad \text{for} \quad \Delta\hat{u}<0\,.
\end{split}\nonumber
\end{equation}
This indicates that the trajectories with the initial conditions $\Delta\hat{u}>0$ correspond to contracting cosmological models while with the initial conditions $\Delta\hat{u}<0$ describe expanding cosmologies. 

Figure~\ref{fig:2} represents the phase space diagrams of evolutional paths of the dynamical system \eqref{eq:dyn_sys} for the special case $\xi=\frac{1}{4}$ in the $(v\,,\hat{u})$ and the original $(v\,,u)$ dynamical variables compactified with a circle at infinity of phase space. In the top panel we present phase space diagrams in $(v\,,\hat{u})$ variables where the arrows indicate direction of time $\eta$ related with the scale factor by \eqref{eq:time_spec}. We see that in the vicinity of the static Einstein universe represented by the unstable critical point $E$ trajectories with initial value $\Delta\hat{u}>0$ correspond to initially contracting cosmological solutions and at some time $\eta$ the $\hat{u}(\eta)$ variable changes the sing which corresponds to the transition for contracting to expanding phase of the evolution. The trajectories in the vicinity of the critical point $E$ with $\Delta\hat{u}<0$ initially correspond to the expanding models. In the bottom panel we present evolutional paths for the system in original $(v\,,u)$ dynamical variables. The arrows indicate direction of the growing scale factor $a$ and now the critical point corresponding the the static Einstein universe is located at infinity $u\to\pm\infty$, where for $u\to+\infty$ it corresponds to a stable critical point while for $u\to-\infty$ to an unstable critical point. 

The evolutional paths for the canonical scalar field represented on the left panel in figure~\ref{fig:2} starting form the unstable critical point $E$ corresponding to the Einstein static universe, continue toward the stable critical point $dS_{+}$ corresponding to the de Sitter state. There is an open and dense set of initial conditions giving rise to this non-singular and generic type of evolution. Additionally, in the phase space, there is no non-hyperbolic critical points which indicates that the dynamical system is structurally stable.

\section{Non-minimal coupling and the conformal invariance}

The Klein-Gordon equation for the scalar field with a monomial potential function $U(\phi)=U_{0}\phi^{n}$ is in the following form
$$
\Box\phi-\xi R\phi-\ve\,n \,U_{0}\phi^{n-1}=0\,,
$$
which in general is not conformally invariant. However, using appropriate conformal or Weyl transformation we can find constraints on the parameters for which the matter sector of gravitational theory is conformally invariant. Working in $D\ge2$ space-time dimensions and using the following point dependent rescaling of the metric tensor and the scalar field
$$
\tilde{g}_{\mu\nu}=\Omega^{2}g_{\mu\nu}\,,\quad \tilde{\phi}=\Omega^{-\frac{D-2}{2}}\phi\,,
$$
where $\Omega$ is a regular, nowhere vanishing function on a smooth manifold we obtain
\begin{equation}
\begin{split}
\tilde{\Box}\tilde{\phi}-\xi \tilde{R}\tilde{\phi}-\ve\,n\, U_{0}\tilde{\phi}^{n-1}{\color{white}\Big)}& =\\\Omega^{-\frac{D+2}{2}}\Big(\Box\phi-\xi R\phi-\ve\,n\,U_{0}\phi^{n-1}\Big)&=0\,,
\end{split}\nonumber
\end{equation}
and this equation holds iff the parameters are 
$$
\xi=\xi_{\textrm{conf}}=\frac{1}{4}\frac{D-2}{D-1}\,,\quad n=n_{\textrm{conf}}=\frac{2D}{D-2}\,.
$$
Hence, we obtain the following discrete set of theoretically motivated values of the non-minimal coupling constant and the exponent of a monomial scalar field potential function suggested by the conformal invariance condition in $D\ge2$ space-time dimensions
\begin{equation}
\begin{split}
\Big\{(D,\xi,n)\Big\}=\Big\{&\left(2,0,\infty\right),\left(3,\frac{1}{8},6\right),\left(4,\frac{1}{6},4\right),\\& \left(5,\frac{3}{16},\frac{10}{3}\right),\dots,\left(\infty,\frac{1}{4},2\right)\Big\}\,.
\end{split}\nonumber
\end{equation}
We can observe that those theoretically motivated values are located within observationally obtained intervals \cite{Hrycyna:2015vvs} and the constraints from quantum cosmology consideration \cite{Wang:2019spw}. Values of the non-minimal coupling constant $\frac{3}{16}<\xi<\frac{1}{4}$ lead to extra-dimensional conformally coupled scalar field theory.    Additionally, we have to note that for a quadratic potential function investigated in this paper the value of the non-minimal coupling constant $\xi=\frac{1}{4}$ may suggest existence of the additional fundamental conformal invariance in the matter sector of the theory \cite{Nakayama:2013is,Bars:2013yba,Englert:1975wj,Englert:1976ep,tHooft:2011aa,tHooft:2014daa,tHooft:2015} and necessity of postulating large number or even an infinite number of extra dimensions \cite{Strominger:1981jg,Deser:1998ed,Sloan:2016kbc}.

\section{Conclusions}
We have investigated dynamics of a flat FRW cosmological model filled with the non-minimally coupled scalar field. We have found that for generic scalar field potential functions which asymptotically tend to a quadratic form at infinite values of the scalar field there is an unstable critical point corresponding to two interesting physical states. For the non-minimal coupling constant $\frac{3}{16}<\xi<\frac{1}{4}$  we found the unstable de Sitter state, additionally free from a parallelly propagated singularity for $\frac{5}{24}\le\xi<\frac{1}{4}$, as an initial state for the evolution of universe with vanishing value of the scale factor but finite value of the energy density. The energy density at this de Sitter state depends on the non-minimal coupling constant and can be smaller that the Planck energy density. Additionally, we have shown that for special value of the non-minimal coupling constant $\xi=\frac{1}{4}$ the initial state for the evolution of universe is an unstable Einstein static universe filled with effective matter in the form of Zeldovich stiff matter.

We have shown that for those two types of initial state for universe there is an open and dense set of initial conditions leading to non-singular evolution. In the first case for $\frac{3}{16}<\xi<\frac{1}{4}$ and the canonical $\ve=+1$ scalar field (see Fig.~\ref{fig:1}) there are evolutional paths connecting the initial unstable de Sitter state with the stable final de Sitter state. In the second case for $\xi=\frac{1}{4}$ and also for the canonical scalar field $\ve=+1$ (see Fig.~\ref{fig:2}) there are evolutional paths starting from the unstable Einstein static state which asymptotically tend to the stable de Sitter state. Note that for the phantom scalar field $\ve=-1$ we are unable to connect the unstable de Sitter state (or the unstable Einstein state) with the sable de Sitter state. 

Using dynamical systems methods we were able to directly compare dynamics of the model in original Jordan frame and conformally transformed Einstein frame. We have shown that both descriptions are physically nonequivalent since the initial unstable de Sitter state in the Jordan frame is transformed in to the stable Einstein static state in the Einstein frame.

The non-singular evolutional paths found in this paper, in contrast to the seminal Starobinsky type of evolution \cite{Starobinsky:1980te,Starobinsky:1978}, are generic in the sense that they do not depend on specific choice of initial conditions in the phase space nor were carefully designated \cite{Mukhanov:1991zn,Brandenberger:1993ef}. Moreover the phase space of the model is organised only by hyperbolic critical points with no trajectories connecting saddle type critical points, which leads to structurally stable dynamics \cite{Andronov:1937, Nielsen:1991aj, Thom:book}.


\bibliographystyle{elsarticle-num}
\bibliography{../bib/moje,../bib/darkenergy,../bib/quintessence,../bib/quartessence,../bib/astro,../bib/dynamics,../bib/standard,../bib/inflation,../bib/confinv,../bib/sm_nmc,../bib/JvsE,../bib/extra_D,../bib/singularities}

\begin{thebibliography}{10}
\expandafter\ifx\csname url\endcsname\relax
  \def\url#1{\texttt{#1}}\fi
\expandafter\ifx\csname urlprefix\endcsname\relax\def\urlprefix{URL }\fi
\expandafter\ifx\csname href\endcsname\relax
  \def\href#1#2{#2} \def\path#1{#1}\fi

\bibitem{Riess:1998cb}
A.~G. Riess, A.~V. Filippenko, P.~Challis, A.~Clocchiattia, A.~Diercks, P.~M.
  Garnavich, R.~L. Gilliland, C.~J. Hogan, S.~Jha, R.~P. Kirshner,
  B.~Leibundgut, M.~M. Phillips, D.~Reiss, B.~P. Schmidt, R.~A. Schommer, R.~C.
  Smith, J.~Spyromilio, C.~Stubbs, N.~B. Suntzeff, J.~Tonry, Observational
  evidence from supernovae for an accelerating universe and a cosmological
  constant, Astron. J. 116 (1998) 1009--1038.
\newblock \href {http://arxiv.org/abs/astro-ph/9805201}
  {\path{arXiv:astro-ph/9805201}}, \href {https://doi.org/10.1086/300499}
  {\path{doi:10.1086/300499}}.

\bibitem{Perlmutter:1998np}
S.~Perlmutter, G.~Aldering, G.~Goldhaber, R.~Knop, P.~Nugent, P.~Castro,
  S.~Deustua, S.~Fabbro, A.~Goobar, D.~Groom, I.~M. Hook, A.~Kim, M.~Kim,
  J.~Lee, N.~Nunes, C.~P. R.~Pain, R.~Quimby, C.~Lidman, R.~Ellis, M.~Irwin,
  R.~McMahon, P.~Ruiz-Lapuente, N.~Walton, B.~Schaefer, B.~Boyle,
  A.~Filippenko, T.~Matheson, A.~Fruchter, N.~Panagia, H.~Newberg, W.~Couch,
  Measurements of omega and lambda from 42 high-redshift supernovae, Astrophys.
  J. 517 (1999) 565--586.
\newblock \href {http://arxiv.org/abs/astro-ph/9812133}
  {\path{arXiv:astro-ph/9812133}}, \href {https://doi.org/10.1086/307221}
  {\path{doi:10.1086/307221}}.

\bibitem{Copeland:2006wr}
E.~J. Copeland, M.~Sami, S.~Tsujikawa, {Dynamics of dark energy},
  Int.~J.~Mod.~Phys. D15 (2006) 1753--1936.
\newblock \href {http://arxiv.org/abs/hep-th/0603057}
  {\path{arXiv:hep-th/0603057}}, \href
  {https://doi.org/10.1142/S021827180600942X}
  {\path{doi:10.1142/S021827180600942X}}.

\bibitem{Bahamonde:2017ize}
S.~Bahamonde, C.~G. Böhmer, S.~Carloni, E.~J. Copeland, W.~Fang, N.~Tamanini,
  {Dynamical systems applied to cosmology: dark energy and modified gravity},
  Phys. Rept. 775-777 (2018) 1--122.
\newblock \href {http://arxiv.org/abs/1712.03107} {\path{arXiv:1712.03107}},
  \href {https://doi.org/10.1016/j.physrep.2018.09.001}
  {\path{doi:10.1016/j.physrep.2018.09.001}}.

\bibitem{Peebles:1987ek}
P.~J.~E. Peebles, B.~Ratra, {Cosmology with a Time Variable Cosmological
  Constant}, Astrophys. J. Lett. 325 (1988) L17.
\newblock \href {https://doi.org/10.1086/185100} {\path{doi:10.1086/185100}}.

\bibitem{Ratra:1987rm}
B.~Ratra, P.~J.~E. Peebles, {Cosmological consequences of a rolling homogeneous
  scalar field}, Phys.~Rev. D37 (1988) 3406.
\newblock \href {https://doi.org/10.1103/PhysRevD.37.3406}
  {\path{doi:10.1103/PhysRevD.37.3406}}.

\bibitem{Wetterich:1987fm}
C.~Wetterich, {Cosmology and the fate of dilatation symmetry}, Nucl.~Phys. B302
  (1988) 668.
\newblock \href {http://arxiv.org/abs/1711.03844} {\path{arXiv:1711.03844}},
  \href {https://doi.org/10.1016/0550-3213(88)90193-9}
  {\path{doi:10.1016/0550-3213(88)90193-9}}.

\bibitem{Chernikov:1968zm}
N.~A. Chernikov, E.~A. Tagirov, {Quantum theory of scalar fields in de Sitter
  space-time}, Annales Poincare Phys. Theor. A9 (1968) 109.

\bibitem{Callan:1970ze}
C.~G. Callan, Jr., S.~R. Coleman, R.~Jackiw, {A New improved energy-momentum
  tensor}, Annals Phys. 59 (1970) 42--73.
\newblock \href {https://doi.org/10.1016/0003-4916(70)90394-5}
  {\path{doi:10.1016/0003-4916(70)90394-5}}.

\bibitem{Birrell:1979ip}
N.~D. Birrell, P.~C.~W. Davies, {Conformal-symmetry breaking and cosmological
  particle creation in $\lambda\phi^{4}$ theory}, Phys. Rev. D22 (1980) 322.
\newblock \href {https://doi.org/10.1103/PhysRevD.22.322}
  {\path{doi:10.1103/PhysRevD.22.322}}.

\bibitem{Donoghue:1994dn}
J.~F. Donoghue, {General relativity as an effective field theory: The leading
  quantum corrections}, Phys.Rev. D50 (1994) 3874--3888.
\newblock \href {http://arxiv.org/abs/gr-qc/9405057}
  {\path{arXiv:gr-qc/9405057}}, \href
  {https://doi.org/10.1103/PhysRevD.50.3874}
  {\path{doi:10.1103/PhysRevD.50.3874}}.

\bibitem{Birrell:1984ix}
N.~D. Birrell, P.~C.~W. Davies, Quantum Fields in Curved Space, Cambridge
  University Press, Cambridge, 1984.

\bibitem{Parker:book}
L.~E. Parker, D.~J. Toms, {Quantum Field Theory in Curved Spacetime. Quantized
  Fields and Gravity}, Cambridge University Press, Cambridge, 2009.

\bibitem{Maeda:1985bq}
K.-i. Maeda, {Stability and attractor in a higher-dimensional cosmology. I},
  Class. Quant. Grav. 3 (1986) 233.
\newblock \href {https://doi.org/10.1088/0264-9381/3/2/017}
  {\path{doi:10.1088/0264-9381/3/2/017}}.

\bibitem{Accetta:1985du}
F.~S. Accetta, D.~J. Zoller, M.~S. Turner, {Induced Gravity Inflation}, Phys.
  Rev. D31 (1985) 3046.
\newblock \href {https://doi.org/10.1103/PhysRevD.31.3046}
  {\path{doi:10.1103/PhysRevD.31.3046}}.

\bibitem{Muta:1991mw}
T.~Muta, S.~D. Odintsov, {Model dependence of the nonminimal scalar graviton
  effective coupling constant in curved space-time}, Mod. Phys. Lett. A 6
  (1991) 3641--3646.
\newblock \href {https://doi.org/10.1142/S0217732391004206}
  {\path{doi:10.1142/S0217732391004206}}.

\bibitem{Buchbinder:1992rb}
I.~L. Buchbinder, S.~D. Odintsov, I.~L. Shapiro, {Effective action in quantum
  gravity}, Taylor \& Francis, New York\, London, 1992.

\bibitem{Atkins:2010eq}
M.~Atkins, X.~Calmet, {On the unitarity of linearized General Relativity
  coupled to matter}, Phys. Lett. B695 (2011) 298--302.
\newblock \href {http://arxiv.org/abs/1002.0003} {\path{arXiv:1002.0003}},
  \href {https://doi.org/10.1016/j.physletb.2010.10.049}
  {\path{doi:10.1016/j.physletb.2010.10.049}}.

\bibitem{Atkins:2010re}
M.~Atkins, X.~Calmet, {Unitarity bounds on low scale quantum gravity}, Eur.
  Phys. J. C70 (2010) 381--388.
\newblock \href {http://arxiv.org/abs/1005.1075} {\path{arXiv:1005.1075}},
  \href {https://doi.org/10.1140/epjc/s10052-010-1476-2}
  {\path{doi:10.1140/epjc/s10052-010-1476-2}}.

\bibitem{Luo:2005ra}
M.-X. Luo, Q.-P. Su, {Fitting Non-Minimally Coupled Scalar Models to Gold SnIa
  Dataset}, Phys.~Lett. B626 (2005) 7--14.
\newblock \href {http://arxiv.org/abs/astro-ph/0506093}
  {\path{arXiv:astro-ph/0506093}}, \href
  {https://doi.org/10.1016/j.physletb.2005.08.050}
  {\path{doi:10.1016/j.physletb.2005.08.050}}.

\bibitem{Nozari:2007eq}
K.~Nozari, S.~D. Sadatian, {Non-Minimal Inflation after WMAP3},
  Mod.~Phys.~Lett. A23 (2008) 2933--2945.
\newblock \href {http://arxiv.org/abs/0710.0058} {\path{arXiv:0710.0058}},
  \href {https://doi.org/10.1142/S0217732308026698}
  {\path{doi:10.1142/S0217732308026698}}.

\bibitem{Szydlowski:2008zza}
M.~Szydlowski, O.~Hrycyna, A.~Kurek, {Coupling constant constraints in a
  nonminimally coupled phantom cosmology}, Phys. Rev. D77 (2008) 027302.
\newblock \href {http://arxiv.org/abs/0710.0366} {\path{arXiv:0710.0366}},
  \href {https://doi.org/10.1103/PhysRevD.77.027302}
  {\path{doi:10.1103/PhysRevD.77.027302}}.

\bibitem{Atkins:2012yn}
M.~Atkins, X.~Calmet, {Bounds on the Nonminimal Coupling of the Higgs Boson to
  Gravity}, Phys.~Rev.~Lett. 110 (2013) 051301.
\newblock \href {http://arxiv.org/abs/1211.0281} {\path{arXiv:1211.0281}},
  \href {https://doi.org/10.1103/PhysRevLett.110.051301}
  {\path{doi:10.1103/PhysRevLett.110.051301}}.

\bibitem{Hrycyna:2015vvs}
O.~Hrycyna, {What $\xi$? Cosmological constraints on the non-minimal coupling
  constant}, Phys. Lett. B768 (2017) 218--227.
\newblock \href {http://arxiv.org/abs/1511.08736} {\path{arXiv:1511.08736}},
  \href {https://doi.org/10.1016/j.physletb.2017.02.062}
  {\path{doi:10.1016/j.physletb.2017.02.062}}.

\bibitem{Spokoiny:1984bd}
B.~L. Spokoiny, {Inflation and generation of perturbations in broken-symmetric
  theory of gravity}, Phys.~Lett. B147 (1984) 39--43.
\newblock \href {https://doi.org/10.1016/0370-2693(84)90587-2}
  {\path{doi:10.1016/0370-2693(84)90587-2}}.

\bibitem{Belinsky:1985zd}
V.~A. Belinsky, I.~M. Khalatnikov, L.~P. Grishchuk, Y.~B. Zeldovich,
  {Inflationary stages in cosmological models with a scalar field}, Phys.~Lett.
  B155 (1985) 232--236.
\newblock \href {https://doi.org/10.1016/0370-2693(85)90644-6}
  {\path{doi:10.1016/0370-2693(85)90644-6}}.

\bibitem{Amendola:1990nn}
L.~Amendola, M.~Litterio, F.~Occhionero, {The Phase space view of inflation. 1:
  The nonminimally coupled scalar field}, Int.~J.~Mod.~Phys. A5 (1990)
  3861--3886.
\newblock \href {https://doi.org/10.1142/S0217751X90001653}
  {\path{doi:10.1142/S0217751X90001653}}.

\bibitem{Faraoni:1996rf}
V.~Faraoni, {Non-minimal coupling of the scalar field and inflation},
  Phys.~Rev. D53 (1996) 6813--6821.
\newblock \href {http://arxiv.org/abs/astro-ph/9602111}
  {\path{arXiv:astro-ph/9602111}}, \href
  {https://doi.org/10.1103/PhysRevD.53.6813}
  {\path{doi:10.1103/PhysRevD.53.6813}}.

\bibitem{Barvinsky:2008ia}
A.~O. Barvinsky, A.~Y. Kamenshchik, A.~A. Starobinsky, {Inflation scenario via
  the Standard Model Higgs boson and LHC}, JCAP 11 (2008) 021.
\newblock \href {http://arxiv.org/abs/0809.2104} {\path{arXiv:0809.2104}},
  \href {https://doi.org/10.1088/1475-7516/2008/11/021}
  {\path{doi:10.1088/1475-7516/2008/11/021}}.

\bibitem{Setare:2008pc}
M.~R. Setare, E.~N. Saridakis, {Non-minimally coupled canonical, phantom and
  quintom models of holographic dark energy}, Phys.~Lett. B671 (2009) 331--338.
\newblock \href {http://arxiv.org/abs/0810.0645} {\path{arXiv:0810.0645}},
  \href {https://doi.org/10.1016/j.physletb.2008.12.026}
  {\path{doi:10.1016/j.physletb.2008.12.026}}.

\bibitem{Uzan:1999ch}
J.-P. Uzan, {Cosmological scaling solutions of non-minimally coupled scalar
  fields}, Phys.~Rev. D59 (1999) 123510.
\newblock \href {http://arxiv.org/abs/gr-qc/9903004}
  {\path{arXiv:gr-qc/9903004}}, \href
  {https://doi.org/10.1103/PhysRevD.59.123510}
  {\path{doi:10.1103/PhysRevD.59.123510}}.

\bibitem{Amendola:1999qq}
L.~Amendola, {Scaling solutions in general non-minimal coupling theories},
  Phys.~Rev. D60 (1999) 043501.
\newblock \href {http://arxiv.org/abs/astro-ph/9904120}
  {\path{arXiv:astro-ph/9904120}}, \href
  {https://doi.org/10.1103/PhysRevD.60.043501}
  {\path{doi:10.1103/PhysRevD.60.043501}}.

\bibitem{Holden:1999hm}
D.~J. Holden, D.~Wands, {Self-similar cosmological solutions with a
  non-minimally coupled scalar field}, Phys.~Rev. D61 (2000) 043506.
\newblock \href {http://arxiv.org/abs/gr-qc/9908026}
  {\path{arXiv:gr-qc/9908026}}, \href
  {https://doi.org/10.1103/PhysRevD.61.043506}
  {\path{doi:10.1103/PhysRevD.61.043506}}.

\bibitem{Gannouji:2006jm}
R.~Gannouji, D.~Polarski, A.~Ranquet, A.~A. Starobinsky, {Scalar-tensor models
  of normal and phantom dark energy}, JCAP 09 (2006) 016.
\newblock \href {http://arxiv.org/abs/astro-ph/0606287}
  {\path{arXiv:astro-ph/0606287}}, \href
  {https://doi.org/10.1088/1475-7516/2006/09/016}
  {\path{doi:10.1088/1475-7516/2006/09/016}}.

\bibitem{Carloni:2007eu}
S.~Carloni, S.~Capozziello, J.~A. Leach, P.~K.~S. Dunsby, {Cosmological
  dynamics of scalar-tensor gravity}, Class.~Quant.~Grav. 25 (2008) 035008.
\newblock \href {http://arxiv.org/abs/gr-qc/0701009}
  {\path{arXiv:gr-qc/0701009}}, \href
  {https://doi.org/10.1088/0264-9381/25/3/035008}
  {\path{doi:10.1088/0264-9381/25/3/035008}}.

\bibitem{Bezrukov:2007ep}
F.~L. Bezrukov, M.~Shaposhnikov, {The Standard Model Higgs boson as the
  inflaton}, Phys.~Lett. B659 (2008) 703--706.
\newblock \href {http://arxiv.org/abs/0710.3755} {\path{arXiv:0710.3755}},
  \href {https://doi.org/10.1016/j.physletb.2007.11.072}
  {\path{doi:10.1016/j.physletb.2007.11.072}}.

\bibitem{Kamenshchik:1995ib}
A.~Y. Kamenshchik, I.~M. Khalatnikov, A.~V. Toporensky, {Nonminimally coupled
  complex scalar field in classical and quantum cosmology}, Phys.~Lett. B357
  (1995) 36--42.
\newblock \href {http://arxiv.org/abs/gr-qc/9508034}
  {\path{arXiv:gr-qc/9508034}}, \href
  {https://doi.org/10.1016/0370-2693(95)00834-8}
  {\path{doi:10.1016/0370-2693(95)00834-8}}.

\bibitem{DeSimone:2008ei}
A.~De~Simone, M.~P. Hertzberg, F.~Wilczek, {Running Inflation in the Standard
  Model}, Phys.~Lett. B678 (2009) 1--8.
\newblock \href {http://arxiv.org/abs/0812.4946} {\path{arXiv:0812.4946}},
  \href {https://doi.org/10.1016/j.physletb.2009.05.054}
  {\path{doi:10.1016/j.physletb.2009.05.054}}.

\bibitem{Bezrukov:2008ej}
F.~L. Bezrukov, A.~Magnin, M.~Shaposhnikov, {Standard Model Higgs boson mass
  from inflation}, Phys.~Lett. B675 (2009) 88--92.
\newblock \href {http://arxiv.org/abs/0812.4950} {\path{arXiv:0812.4950}},
  \href {https://doi.org/10.1016/j.physletb.2009.03.035}
  {\path{doi:10.1016/j.physletb.2009.03.035}}.

\bibitem{Barvinsky:2009fy}
A.~O. Barvinsky, A.~Y. Kamenshchik, C.~Kiefer, A.~A. Starobinsky,
  C.~Steinwachs, {Asymptotic freedom in inflationary cosmology with a
  non-minimally coupled Higgs field}, JCAP 12 (2009) 003.
\newblock \href {http://arxiv.org/abs/0904.1698} {\path{arXiv:0904.1698}},
  \href {https://doi.org/10.1088/1475-7516/2009/12/003}
  {\path{doi:10.1088/1475-7516/2009/12/003}}.

\bibitem{Clark:2009dc}
T.~E. Clark, B.~Liu, S.~T. Love, T.~ter Veldhuis, {The Standard Model Higgs
  Boson-Inflaton and Dark Matter}, Phys.~Rev. D80 (2009) 075019.
\newblock \href {http://arxiv.org/abs/0906.5595} {\path{arXiv:0906.5595}},
  \href {https://doi.org/10.1103/PhysRevD.80.075019}
  {\path{doi:10.1103/PhysRevD.80.075019}}.

\bibitem{Hrycyna:2010yv}
O.~Hrycyna, M.~Szydlowski, {Uniting cosmological epochs through the twister
  solution in cosmology with non-minimal coupling}, JCAP 12 (2010) 016.
\newblock \href {http://arxiv.org/abs/1008.1432} {\path{arXiv:1008.1432}},
  \href {https://doi.org/10.1088/1475-7516/2010/12/016}
  {\path{doi:10.1088/1475-7516/2010/12/016}}.

\bibitem{Hrycyna:2015eta}
O.~Hrycyna, M.~Szydlowski, {Cosmological dynamics with non-minimally coupled
  scalar field and a constant potential function}, JCAP 11 (2015) 013.
\newblock \href {http://arxiv.org/abs/1506.03429} {\path{arXiv:1506.03429}},
  \href {https://doi.org/10.1088/1475-7516/2015/11/013}
  {\path{doi:10.1088/1475-7516/2015/11/013}}.

\bibitem{Hrycyna:2020jmw}
O.~Hrycyna, {The non-minimal coupling constant and the primordial de Sitter
  state}, Eur. Phys. J. C 80 (2020) 817.
\newblock \href {http://arxiv.org/abs/2008.00943} {\path{arXiv:2008.00943}},
  \href {https://doi.org/10.1140/epjc/s10052-020-8397-5}
  {\path{doi:10.1140/epjc/s10052-020-8397-5}}.

\bibitem{Kerachian:2019tar}
M.~Kerachian, G.~Acquaviva, G.~Lukes-Gerakopoulos, {Classes of nonminimally
  coupled scalar fields in spatially curved FRW spacetimes}, Phys. Rev. D
  99~(12) (2019) 123516.
\newblock \href {http://arxiv.org/abs/1905.08512} {\path{arXiv:1905.08512}},
  \href {https://doi.org/10.1103/PhysRevD.99.123516}
  {\path{doi:10.1103/PhysRevD.99.123516}}.

\bibitem{Jarv:2021qpp}
L.~J\"arv, A.~Toporensky, {Global portraits of nonminimal inflation} (4 2021).
\newblock \href {http://arxiv.org/abs/2104.10183} {\path{arXiv:2104.10183}}.

\bibitem{Hrycyna:2007mq}
O.~Hrycyna, M.~Szydlowski, {Route to Lambda in conformally coupled phantom
  cosmology}, Phys.~Lett. B651 (2007) 8--14.
\newblock \href {http://arxiv.org/abs/0704.1651} {\path{arXiv:0704.1651}},
  \href {https://doi.org/10.1016/j.physletb.2007.05.057}
  {\path{doi:10.1016/j.physletb.2007.05.057}}.

\bibitem{Hrycyna:2008gk}
O.~Hrycyna, M.~Szydlowski, {Non-minimally coupled scalar field cosmology on the
  phase plane}, JCAP 04 (2009) 026.
\newblock \href {http://arxiv.org/abs/0812.5096} {\path{arXiv:0812.5096}},
  \href {https://doi.org/10.1088/1475-7516/2009/04/026}
  {\path{doi:10.1088/1475-7516/2009/04/026}}.

\bibitem{Felder:2002jk}
G.~N. Felder, A.~V. Frolov, L.~Kofman, A.~D. Linde, {Cosmology with negative
  potentials}, Phys. Rev. D 66 (2002) 023507.
\newblock \href {http://arxiv.org/abs/hep-th/0202017}
  {\path{arXiv:hep-th/0202017}}, \href
  {https://doi.org/10.1103/PhysRevD.66.023507}
  {\path{doi:10.1103/PhysRevD.66.023507}}.

\bibitem{Boisseau:2015hqa}
B.~Boisseau, H.~Giacomini, D.~Polarski, A.~A. Starobinsky, {Bouncing Universes
  in Scalar--Tensor Gravity Models admitting Negative Potentials}, JCAP 07
  (2015) 002.
\newblock \href {http://arxiv.org/abs/1504.07927} {\path{arXiv:1504.07927}},
  \href {https://doi.org/10.1088/1475-7516/2015/07/002}
  {\path{doi:10.1088/1475-7516/2015/07/002}}.

\bibitem{Linde:2001ae}
A.~D. Linde, {Fast-Roll Inflation}, JHEP 11 (2001) 052.
\newblock \href {http://arxiv.org/abs/hep-th/0110195}
  {\path{arXiv:hep-th/0110195}}, \href
  {https://doi.org/10.1088/1126-6708/2001/11/052}
  {\path{doi:10.1088/1126-6708/2001/11/052}}.

\bibitem{Kofman:2007tr}
L.~Kofman, S.~Mukohyama, {Rapid roll Inflation with Conformal Coupling}, Phys.
  Rev. D77 (2008) 043519.
\newblock \href {http://arxiv.org/abs/0709.1952} {\path{arXiv:0709.1952}},
  \href {https://doi.org/10.1103/PhysRevD.77.043519}
  {\path{doi:10.1103/PhysRevD.77.043519}}.

\bibitem{Chiba:2008ia}
T.~Chiba, M.~Yamaguchi, {Extended Slow-Roll Conditions and Rapid-Roll
  Conditions}, JCAP 10 (2008) 021.
\newblock \href {http://arxiv.org/abs/0807.4965} {\path{arXiv:0807.4965}},
  \href {https://doi.org/10.1088/1475-7516/2008/10/021}
  {\path{doi:10.1088/1475-7516/2008/10/021}}.

\bibitem{Yoshida:2018ndv}
D.~Yoshida, J.~Quintin, {Maximal extensions and singularities in inflationary
  spacetimes}, Class. Quant. Grav. 35~(15) (2018) 155019.
\newblock \href {http://arxiv.org/abs/1803.07085} {\path{arXiv:1803.07085}},
  \href {https://doi.org/10.1088/1361-6382/aacf4b}
  {\path{doi:10.1088/1361-6382/aacf4b}}.

\bibitem{Nomura:2021lzz}
K.~Nomura, D.~Yoshida, {Past extendibility and initial singularity in
  Friedmann-Lema\^\i{}tre-Robertson-Walker and Bianchi I spacetimes} (5 2021).
\newblock \href {http://arxiv.org/abs/2105.05642} {\path{arXiv:2105.05642}}.

\bibitem{Cline:2003gs}
J.~M. Cline, S.~Jeon, G.~D. Moore, {The Phantom menaced: Constraints on
  low-energy effective ghosts}, Phys. Rev. D 70 (2004) 043543.
\newblock \href {http://arxiv.org/abs/hep-ph/0311312}
  {\path{arXiv:hep-ph/0311312}}, \href
  {https://doi.org/10.1103/PhysRevD.70.043543}
  {\path{doi:10.1103/PhysRevD.70.043543}}.

\bibitem{Ellis:2002we}
G.~F. Ellis, R.~Maartens, {The emergent universe: Inflationary cosmology with
  no singularity}, Class. Quant. Grav. 21 (2004) 223--232.
\newblock \href {http://arxiv.org/abs/gr-qc/0211082}
  {\path{arXiv:gr-qc/0211082}}, \href
  {https://doi.org/10.1088/0264-9381/21/1/015}
  {\path{doi:10.1088/0264-9381/21/1/015}}.

\bibitem{Ellis:2003qz}
G.~F. Ellis, J.~Murugan, C.~G. Tsagas, {The Emergent universe: An Explicit
  construction}, Class. Quant. Grav. 21~(1) (2004) 233--250.
\newblock \href {http://arxiv.org/abs/gr-qc/0307112}
  {\path{arXiv:gr-qc/0307112}}, \href
  {https://doi.org/10.1088/0264-9381/21/1/016}
  {\path{doi:10.1088/0264-9381/21/1/016}}.

\bibitem{Barrow:2003ni}
J.~D. Barrow, G.~F. Ellis, R.~Maartens, C.~G. Tsagas, {On the stability of the
  Einstein static universe}, Class. Quant. Grav. 20 (2003) L155--L164.
\newblock \href {http://arxiv.org/abs/gr-qc/0302094}
  {\path{arXiv:gr-qc/0302094}}, \href
  {https://doi.org/10.1088/0264-9381/20/11/102}
  {\path{doi:10.1088/0264-9381/20/11/102}}.

\bibitem{Zeldovich:1962}
{\relax Ya}.~B. Zeldovich, {The equation of state at ultrahigh densities and
  its relativistic limitations}, Sov.~Phys.~JETP 14 (1962) 1143--1147,
  [Zh.~Eksp.~Teor.~Fiz.~41, 1609 (1961)].

\bibitem{Zeldovich:1972zz}
{\relax Ya}.~B. Zeldovich, {A hypothesis, unifying the structure and the
  entropy of the universe}, Mon.~Not.~Roy.~Astron.~Soc. 160 (1972) 1P--3P.
\newblock \href {https://doi.org/10.1093/mnras/160.1.1P}
  {\path{doi:10.1093/mnras/160.1.1P}}.

\bibitem{Wang:2019spw}
S.-J. Wang, M.~Yamada, A.~Vilenkin, {Constraints on non-minimal coupling from
  quantum cosmology}, JCAP 08 (2019) 025.
\newblock \href {http://arxiv.org/abs/1903.11736} {\path{arXiv:1903.11736}},
  \href {https://doi.org/10.1088/1475-7516/2019/08/025}
  {\path{doi:10.1088/1475-7516/2019/08/025}}.

\bibitem{Nakayama:2013is}
Y.~Nakayama, {Scale invariance vs conformal invariance}, Phys.~Rept. 569 (2015)
  1--93.
\newblock \href {http://arxiv.org/abs/1302.0884} {\path{arXiv:1302.0884}},
  \href {https://doi.org/10.1016/j.physrep.2014.12.003}
  {\path{doi:10.1016/j.physrep.2014.12.003}}.

\bibitem{Bars:2013yba}
I.~Bars, P.~Steinhardt, N.~Turok, {Local Conformal Symmetry in Physics and
  Cosmology}, Phys.~Rev. D89~(4) (2014) 043515.
\newblock \href {http://arxiv.org/abs/1307.1848} {\path{arXiv:1307.1848}},
  \href {https://doi.org/10.1103/PhysRevD.89.043515}
  {\path{doi:10.1103/PhysRevD.89.043515}}.

\bibitem{Englert:1975wj}
F.~Englert, E.~Gunzig, C.~Truffin, P.~Windey, {Conformal Invariant General
  Relativity with Dynamical Symmetry Breakdown}, Phys.~Lett. B57 (1975) 73.
\newblock \href {https://doi.org/10.1016/0370-2693(75)90247-6}
  {\path{doi:10.1016/0370-2693(75)90247-6}}.

\bibitem{Englert:1976ep}
F.~Englert, C.~Truffin, R.~Gastmans, {Conformal Invariance in Quantum Gravity},
  Nucl.~Phys. B117 (1976) 407.
\newblock \href {https://doi.org/10.1016/0550-3213(76)90406-5}
  {\path{doi:10.1016/0550-3213(76)90406-5}}.

\bibitem{tHooft:2011aa}
G.~'t~Hooft, {A class of elementary particle models without any adjustable real
  parameters}, Found.~Phys. 41 (2011) 1829--1856.
\newblock \href {http://arxiv.org/abs/1104.4543} {\path{arXiv:1104.4543}},
  \href {https://doi.org/10.1007/s10701-011-9586-8}
  {\path{doi:10.1007/s10701-011-9586-8}}.

\bibitem{tHooft:2014daa}
G.~'t~Hooft, {Local conformal symmetry: The missing symmetry component for
  space and time}, Int.~J.~Mod.~Phys.~D 24 (2015) 1543001.
\newblock \href {http://arxiv.org/abs/1410.6675} {\path{arXiv:1410.6675}},
  \href {https://doi.org/10.1142/S0218271815430014}
  {\path{doi:10.1142/S0218271815430014}}.

\bibitem{tHooft:2015}
G.~'t~Hooft, {Spontaneous breakdown of local conformal invariance in quantum
  gravity}, in: L.~Baulieu, K.~Benakli, M.~R. Douglas, B.~Mansoulie,
  E.~Rabinovici, L.~F. Cugliandolo (Eds.), {Theoretical Physics to Face the
  Challenge of LHC: Lecture Notes of the Les Houches Summer School: Volume 97,
  August 2011}, Oxford University Press, 2015, pp. 209--253.
\newblock \href {https://doi.org/10.1093/acprof:oso/9780198727965.003.0010}
  {\path{doi:10.1093/acprof:oso/9780198727965.003.0010}}.

\bibitem{Strominger:1981jg}
A.~Strominger, {The Inverse Dimensional Expansion in Quantum Gravity},
  Phys.~Rev. D24 (1981) 3082.
\newblock \href {https://doi.org/10.1103/PhysRevD.24.3082}
  {\path{doi:10.1103/PhysRevD.24.3082}}.

\bibitem{Deser:1998ed}
S.~Deser, {Dimensionally challenged gravities}, in: J.~Renn, L.~Divarci,
  P.~Schr{\"o}ter, A.~Ashtekar, R.~S. Cohen, D.~Howard, S.~Sarkar, A.~Shimony
  (Eds.), Revisiting the Foundations of Relativistic Physics: Festschrift in
  Honor of John Stachel, Vol. 234 of Boston Studies in Philosophy of Science,
  Springer Netherlands, Dordrecht, 2003, pp. 397--401.
\newblock \href {http://arxiv.org/abs/gr-qc/9812013}
  {\path{arXiv:gr-qc/9812013}}, \href
  {https://doi.org/10.1007/978-94-010-0111-3_19}
  {\path{doi:10.1007/978-94-010-0111-3_19}}.

\bibitem{Sloan:2016kbc}
D.~Sloan, P.~Ferreira, {The Cosmology of an Infinite Dimensional Universe},
  Phys. Rev. D96~(4) (2017) 043527.
\newblock \href {http://arxiv.org/abs/1612.02853} {\path{arXiv:1612.02853}},
  \href {https://doi.org/10.1103/PhysRevD.96.043527}
  {\path{doi:10.1103/PhysRevD.96.043527}}.

\bibitem{Starobinsky:1980te}
A.~A. Starobinsky, {A new type of isotropic cosmological models without
  singularity}, Phys.~Lett. B91 (1980) 99--102.
\newblock \href {https://doi.org/10.1016/0370-2693(80)90670-X}
  {\path{doi:10.1016/0370-2693(80)90670-X}}.

\bibitem{Starobinsky:1978}
A.~A. Starobinsky, {On a nonsingular isotropic cosmological model},
  Sov.~Astro.~Lett. 4 (1978) 82--84, [Pisma Astron.~Zh.~4, 155-159 (1978)].

\bibitem{Mukhanov:1991zn}
V.~F. Mukhanov, R.~H. Brandenberger, {A Nonsingular universe}, Phys.~Rev.~Lett.
  68 (1992) 1969--1972.
\newblock \href {https://doi.org/10.1103/PhysRevLett.68.1969}
  {\path{doi:10.1103/PhysRevLett.68.1969}}.

\bibitem{Brandenberger:1993ef}
R.~H. Brandenberger, V.~F. Mukhanov, A.~Sornborger, {Cosmological theory
  without singularities}, Phys.~Rev. D48 (1993) 1629--1642.
\newblock \href {http://arxiv.org/abs/gr-qc/9303001}
  {\path{arXiv:gr-qc/9303001}}, \href
  {https://doi.org/10.1103/PhysRevD.48.1629}
  {\path{doi:10.1103/PhysRevD.48.1629}}.

\bibitem{Andronov:1937}
A.~A. Andronov, L.~S. Pontryagin, Grubyye sistemy, Dokl. Akad. Nauk SSSR 14
  (1937) 247--251, [english translation : Lev Semenovich Pontryagin, in Russian
  Mathematicians in the 20th Century, eds. Ya. G. Sinai, World Scientific
  Publishing Co. Pte. Ltd, 2003, pp. 345-366].
\newblock \href {https://doi.org/10.1142/9789812779212_0015}
  {\path{doi:10.1142/9789812779212_0015}}.

\bibitem{Nielsen:1991aj}
H.~B. Nielsen, {Catastrophe theory programme}, in: C.~D. Froggatt, H.~B.
  Nielsen (Eds.), Origin of Symmetries, World Scientific Publishing Co. Pte.
  Ltd., Singapore, 1991, pp. 566--581, [Dual Strings - Section 6. Catastrophe
  Theory Programme by H. B. Nielsen, in Fundamentals of Quark Models, eds. I.
  M. Barbour and A. T. Davies, Scottish Universities Summer School in Physics
  (1976), pp. 528-543].
\newblock \href {https://doi.org/10.1142/9789814329057_0037}
  {\path{doi:10.1142/9789814329057_0037}}.

\bibitem{Thom:book}
R.~Thom, Structural Stalility and Morphogenesis: An Outline of a General Theory
  of Models, Advanced Book Classics, Westview Press, Reading, Mass., 1989.

\end{thebibliography}

\end{document}